%% file: site_version_principal.tex
\documentclass[preprint,12pt]{elsarticle}

\usepackage{amssymb}
\usepackage{color}

\journal{Experimental Thermal and Fluid Science}

\begin{document}

\begin{frontmatter}



\title{The formation and migration of sand ripples in closed conduits: experiments with turbulent water flows \tnoteref{label_note_copyright} \tnoteref{label_note_doi}}

\tnotetext[label_note_copyright]{\copyright 2016. This manuscript version is made available under the CC-BY-NC-ND 4.0 license http://creativecommons.org/licenses/by-nc-nd/4.0/}

\tnotetext[label_note_doi]{Accepted Manuscript for Experimental Thermal and Fluid Science,  v. 71, p. 95-102, 2016, doi:10.1016/j.expthermflusci.2015.10.017}


\author{Jorge Eduar Cardona Florez}
\ead{jecardonaf@fem.unicamp.br}
\author{Erick de Moraes Franklin\corref{cor1}}
\ead{franklin@fem.unicamp.br}

\cortext[cor1]{Corresponding author}

\address{Faculty of Mechanical Engineering, UNICAMP - University of Campinas\\
Rua Mendeleyev, 200 - Campinas - SP - CEP: 13083-860\\
Brazil}

\begin{abstract}
The transport of solid particles by a fluid flow is frequently found in nature and industry. Some examples are the transport of sand in rivers and hydrocarbon pipelines. When the shear stresses exerted by a fluid flow on a granular bed remain moderate, some grains are set in motion without fluidizing the bed; the moving grains form a layer, known as bed load, that moves while maintaining contact with the fixed part of the bed. Under bed load conditions, the granular bed may become unstable, generating ripples and dunes. Sand ripples are commonly observed in closed conduits and pipes such as in petroleum pipelines, sewer systems, and dredging lines. Although of importance for many scientific domains and industrial applications, the formation of ripples in closed conduits is not well understood, and the problem is still open. This paper presents an experimental study on the formation and migration of sand ripples under a turbulent closed-conduit flow and bed-load conditions. In our experiments, fully-developed turbulent water flows were imposed over a granular bed of known granulometry in a transparent channel, and bed load took place. For different water flow rates and grain diameters, the growth and migration of bedforms were filmed by a high-definition camera, and a numerical code was developed to determine the wavelength and celerity of the bedforms from the acquired images. The obtained results are compared with published stability analyses.
\end{abstract}


\end{frontmatter}


\include{texto_site_version}






\bibliography{references}
\bibliographystyle{elsart-num}







\end{document}

%% file: texto_site_version.tex
\section{Introduction}

When a fluid flows over a granular bed, grains can be entrained owing to momentum transfer from the fluid to the grains. If the shear stresses exerted by the fluid on the granular bed remain moderate, some grains are set in motion without fluidizing the bed; the moving grains form a layer, known as a bed load, that moves while maintaining contact with the fixed part of the bed. Bed load is frequently found in nature and industry. In nature it is present, for example, in the erosion of riverbanks and the formation and migration of dunes in deserts. In industry, it is present in petroleum pipelines, pharmaceutical processes, dredging lines, and sewer systems. Under bed load conditions, the granular bed may become unstable, generating ripples and dunes. In the specific case of oil industry, it is common to observe the formation and migration of ripples inside offshore pipelines conveying oil and sand. These ripples increase pressure losses, raising production costs for petroleum extraction. In addition, their migration along the pipeline causes flow transients. Still concerning offshore activities of the oil industry, the pipelines are subject to movements of sand dunes over the sea bed, which interfere with production activities \cite{Morelissen}. Therefore, better knowledge of bed load and associated instabilities is of importance to understand erosion and deposition in nature, and to control various industrial processes.

The formation and migration of ripples and dunes have been the subject of many studies in previous decades. In the aeolian case, the work of Bagnold \cite{Bagnold_1} was followed by many others that contributed to our understanding of aeolian dunes. Some more recent examples are \cite{Herrmann_Sauermann, Kroy_C, Andreotti_1, Elbelrhiti, Parteli_2, Melo}. In the aquatic case, numerous studies were concerned specifically with the stability of a granular bed sheared by a fluid flow \cite{Kennedy, Reynolds, Engelund_1, Fredsoe_1, Engelund_Fredsoe, Richards, Charru_3, Claudin_Andreotti, Ouriemi_2}. The stability of a granular bed is given by the balance between local erosion and deposition; therefore, the mechanisms creating a phase lag between the shape of the granular bed and the bed-load transport rate must be known. Engelund and Fredsoe (1982) \cite{Engelund_Fredsoe} summarize the stability analyses of most of these papers and show that the fluid flow perturbation, the gravitational field, and the inertia of the grains are directly related to the stability of the bed. Owing to its negative phase lag with respect to the bedform, the fluid flow perturbation caused by the shape of the bed is a destabilizing mechanism \cite{Jackson_Hunt, Hunt_1, Weng}, whereas the local slope of the bed (gravity effects) and the grain inertia (relaxation effects) due to their positive phase lag with respect to the fluid flow perturbation are the stabilizing mechanisms \cite{Valance_Langlois, Charru_3, Franklin_4}.
 
Some stability analyses and experimental studies were carried out for the specific case of turbulent flows of liquids in closed conduits. Kuru et al. (1995) \cite{Kuru} presented a theoretical and experimental study of the initial instabilities of granular beds in pipes. Their experimental test section was a $31.1$-$mm$-diameter pipe, with a length of $7\,m$, and they employed mixtures of water and glycerin as the fluid media and glass beads as the granular medium. The authors presented a linear stability analysis of a clear-layer and suspension-layer cocurrent two-phase flow. They did not consider the bed load in their model, although the grains used and the fluid flow conditions were clearly in the bed-load range. They found that instabilities appear because the shear stresses caused by the clear layer on the interface with the suspension layer are shifted upstream with respect to the undulations of this interface. The wavelength they found overpredicts the experimental measurements, probably because they neglected the bed-load layer and the corresponding effects of relaxation. Their experimental results (performed mainly in the turbulent regime) showed that the initial wavelength scales with the flow rate of the fluid; however, their linear stability analysis was not able to explain the experimental results.

Coleman et al. (2003) \cite{Coleman_1} experimentally studied the instabilities of granular beds in a closed conduit. The experimental test section was a $6$-$m$-long horizontal closed conduit with a rectangular cross section ($300\,mm$ wide by $100\,mm$ high), and they employed water as the fluid medium and glass beads as the granular medium. The fluid flow was in the turbulent regime. They found that the initial instabilities have a well-defined wavelength, which scales with the grain diameter but not with the fluid flow. They also found that the formation of ripples in closed-conduit flows and in subcritical open-channel flows are similar, with roughly the same wavelength for the initial ripples.

Franklin (2008) \cite{Franklin_3} experimentally studied the initial instabilities of different granular beds under turbulent water flows. The experimental test section was a $6$-$m$-long horizontal closed conduit with a rectangular cross section ($120\,mm$ wide by $60\,mm$ high) made of a transparent material. He employed water as the fluid medium and glass and zirconium beads as the granular media. The fluid flow was measured by PIV (Particle Image Velocimetry), and the granular bed evolution was measured by using a high-definition camera.  Franklin (2008) \cite{Franklin_3} found that the initial instabilities have a well-defined wavelength, which scales with both the diameters of the grains and the fluid flow.

Franklin (2010) \cite{Franklin_4} presented a linear stability analysis for the specific case of granular beds sheared by the turbulent boundary layers of liquids. The analysis considered the fluid flow perturbation, local slope, and relaxation effects; however, it neglected the presence of free surfaces. Therefore, the analysis is suitable for bedforms scaling with the low regions of the boundary layer or to closed-conduit flows. Franklin (2010) \cite{Franklin_4} showed that the length scale of the initial bedforms varies with the fluid flow conditions.

Franklin (2011) \cite{Franklin_5} proposed a weakly nonlinear analysis for the stability of granular beds in pipes and closed conduits for the specific case of the turbulent flows of liquids. The author demonstrated that, after the linear growth phase, the bed instabilities saturate in amplitude while maintaining the same wavelength of the linear phase.

This paper presents an experimental study on the formation and migration of ripples in closed conduits under turbulent flow conditions. In our experiments, pressure-driven turbulent water flows were imposed over a granular bed of known granulometry in a transparent channel. For different flow rates and grain diameters, the growth and migration of bedforms were filmed by a high-definition camera. The wavelength and celerity of the bedforms were determined by post-processing the images with a numerical code of our own. The results are compared with published stability analyses.

\section{Experimental device}

\subsection{Experimental setup}

\begin{figure}[!ht]
  \begin{center}
    \includegraphics[width=0.90\columnwidth]{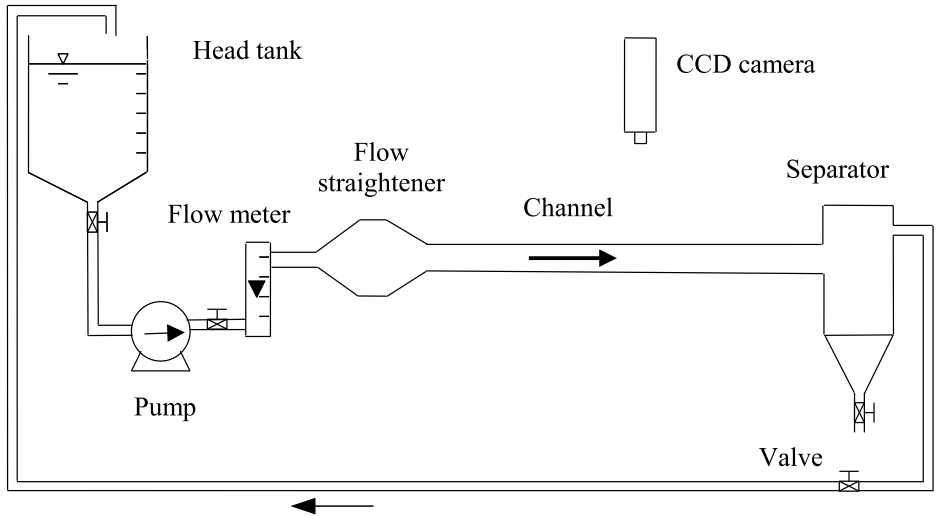}
    \caption{Schematic of the experimental loop.}
    \label{fig:loop}
  \end{center}
\end{figure}

The experimental device consisted of a water reservoir, progressive pump, flow straightener, $5$-$m$-long channel, settling tank, and return line. The flow straightener was a divergent--convergent nozzle filled with $d\,=\,3\,mm$ glass spheres, the function of which was to homogenize the flow profile at the channel inlet. The channel had a rectangular cross section ($160\,mm$ wide by $50\,mm$ high) and was made of a transparent material. Figure \ref{fig:loop} shows a schematic of the experimental loop. The channel test section was $1$ m long and started $40$ hydraulic diameters (3 m) downstream from the channel inlet. A fixed granular bed consisting of glass spheres glued onto the surfaces of PVC plates was inserted into the channel section in which the flow is developed, assuring that the turbulent flow was completely developed in the test section. In the test section, the grains were deposed and formed a loose granular bed having the same thickness ($7\,mm$) as the fixed bed. Glass spheres with a specific mass of $\rho_s\,=\,2500\,kg/m^3$ were employed to form both the fixed and loose granular beds. The grains were classified into three populations, each having their minimum and maximum sizes limited by the use of sieves with distinct meshes: the first had its size ranging from  $d = 212\, \mu m$ to $d = 300\, \mu m$, and it is assumed that the mean diameter was $d_{50}=256\, \mu m$; the second had its size ranging from $d = 300\, \mu m$ to $d = 425\, \mu m$, and it is assumed that $d_{50}=363\, \mu m$; the third had its size ranging from  $d = 500\, \mu m$ to $d = 600\, \mu m$, and it is assumed that $d_{50}= 550\, \mu m$. Prior to each test, the loose granular bed was smoothed and leveled.

The tests were performed at ambient conditions, i.e., an atmospheric pressure of $1\,atm$ and a room temperature of approximately $25^{\circ}C$. The water flowed in a closed loop driven by the pump from the reservoir through the channel and grain separator and back to the reservoir. The water flow rate was controlled by changing the excitation frequency of the pump and was measured with an electromagnetic flow meter. The nominal test flow rates were in the range of $6.6\,m^3/h\,\leq\,Q\,\leq 10.3\,m^3/h$, the cross-section mean velocities were in the range of $0.27\,m/s\,\leq\,U\,\leq\,0.42\,m/s$, and the Reynolds number $Re\,=\, U2H_{gap}/\nu$ was in the range of $2.3\cdot 10^4\,\leq\,Re\,\leq 3.6\cdot 10^4$, where $H_{gap}$ is the distance from the granular bed to the top wall. Franklin et al. (2014) \cite{Franklin_8} measured water flow fields for the same experimental device over similar moving beds. The water flow profiles presented in Franklin et al. (2014) \cite{Franklin_8} were assumed to be valid for the present experiments and are used in this study.

Preset water flows were imposed by the pump, generating a turbulent flow of water over the granular bed to obtain the transport by bed load. Under the effect of the flow, there was the formation and migration of ripples, which were filmed by a high-definition camera. The obtained images were sent to a computer, where they were stored for further processing and analysis.

\begin{figure}[!ht]
  \begin{center}
    \includegraphics[width=0.90\columnwidth]{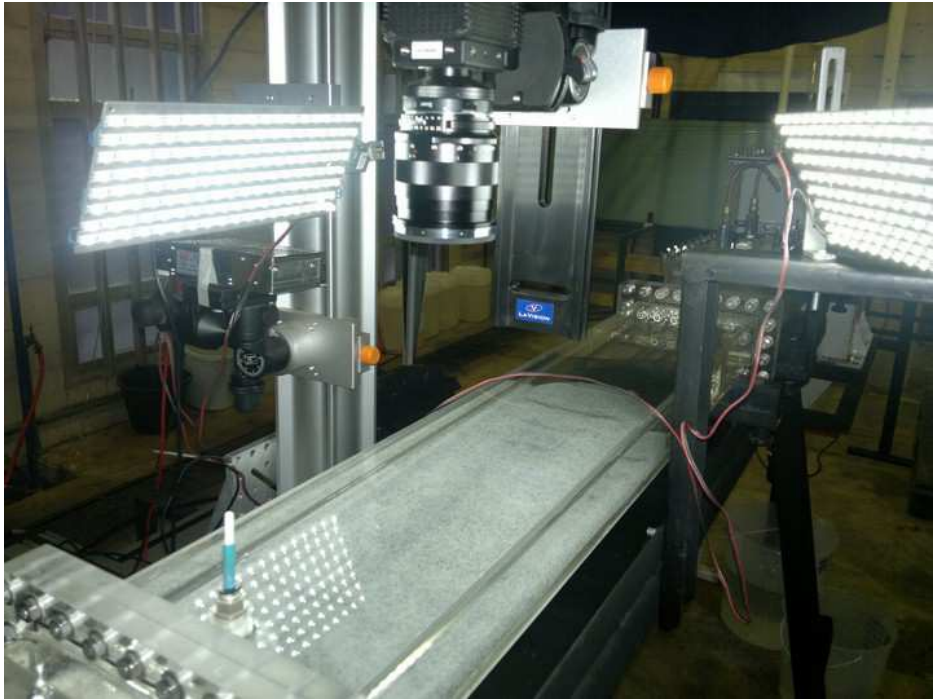}
    \caption{Test section.}
    \label{fig:fotobancada}
  \end{center}
\end{figure}

A high-definition CCD camera with a resolution of $2048\,px\,\times\,2048\,px$ and a 14-bit digital output was employed to film the bed evolution. A computer was used to control the frequencies and exposure times of the high-definition camera and to store the acquired images. In order to provide the necessary light, approximately $200$ LED (Light-Emitting Diode) lamps were attached to plates and branched to a continuous current source. The plates were positioned in order to provide oblique light so that the crests were highlighted. A Makro-Planar lens with a $50$-$mm$ focal distance was used, and the calibration process was performed whenever the camera was displaced, allowing the conversion from pixels to the metric system. The employed fields of view were of $161.6\,mm \,\times\, 161.6\,mm$, $135.3\,mm \,\times\, 135.3\,mm$, and $162.7\,mm \,\times\, 162.7\,mm$ for the $d_{50}=256\, \mu m$, $d_{50}=363\, \mu m$, and $d_{50}= 550\, \mu m$ beds, respectively. Figure \ref{fig:fotobancada} shows the camera and LED plates.

\subsection{Image processing}

A numerical code was developed in order to determine the wavelength and celerity of the bedforms from the high-definition movies. The code is based on the identification of crests and troughs from gray-level differences, and it has two main parts: an image treatment part and a position computation part.

The image treatment comprises three main steps. The first is to homogenize the image brightness by normalizing the mean gray level of each frame. This was necessary because the oblique orientation of the light generated a darker region far from the source. The second step is to find the continuous white regions that are spanwise oriented, corresponding to the crests. This was necessary because the ripples were affected by the vertical channel walls, so that the crest lines (in the spanwise direction) were curved close to vertical walls. Finally, the third step consists of finding the mean position of the center region (far from the vertical walls) of each crest.

The position computation part comprises several steps; it performs the conversion from pixels to millimeters, determines the position of each crest in the image, measures the distances between consecutive crests (wavelength), and computes the celerity from the derivative with respect to time of the crest positions.

\section{Results}

\begin{figure}[!ht]
  \begin{center}
    \includegraphics[width=0.95\columnwidth]{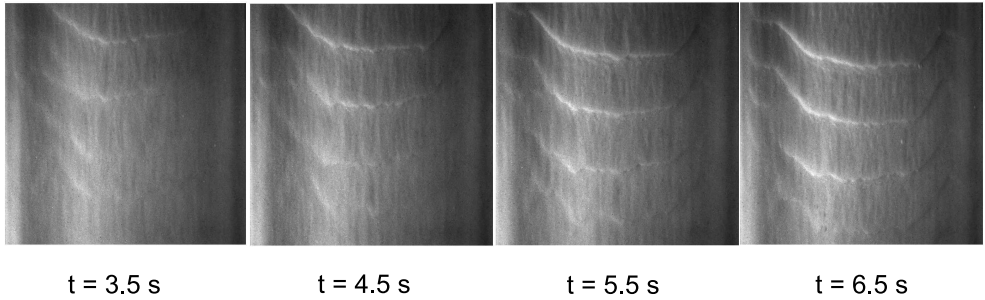}
    \caption{Formation and migration of ripples over a $d_{50}=256\, \mu m$ bed for a water flow rate $Q=8.60\, m^3/h$. The total field of each frame corresponds to $161.6\,mm \,\times\, 161.6\,mm$ and the flow is from top to bottom.}
    \label{photos_rides}
  \end{center}
\end{figure}

For different water flow rates, we filmed the evolution of the granular beds. Figure \ref{photos_rides} shows an example of a frame sequence acquired with the camera--LED system. In this figure, the bed consisted of $d_{50}=256\, \mu m$ grains, and the water flow rate was $Q=8.60\, m^3/h$. The instants corresponding to each frame are presented in the figure. Owing to the oblique light, the ripple crests appear in the figure as brighter regions.

Figure \ref{photos_rides} shows the rapid formation of ripples, on the order of $1\,s$ for this specific granulometry and water flow rate. It also shows the existence of low-velocity regions close to the vertical walls of the channel, curving the crest line of the bedforms. With the exception of these low-velocity regions, the initial bedforms tend to be two-dimensional.

\begin{figure}
   \begin{minipage}[c]{.49\linewidth}
    \begin{center}
     \includegraphics[width=\linewidth]{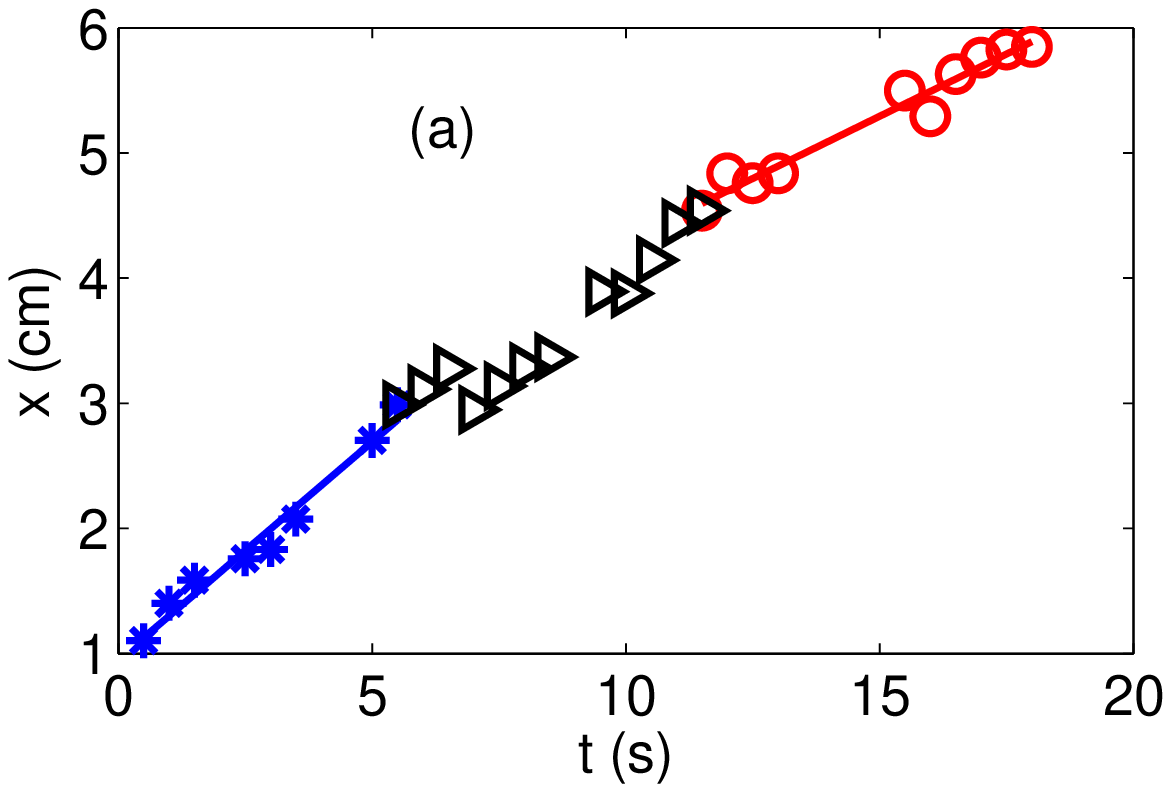}
    \end{center}
   \end{minipage} \hfill
   \begin{minipage}[c]{.49\linewidth}
    \begin{center}
      \includegraphics[width=\linewidth]{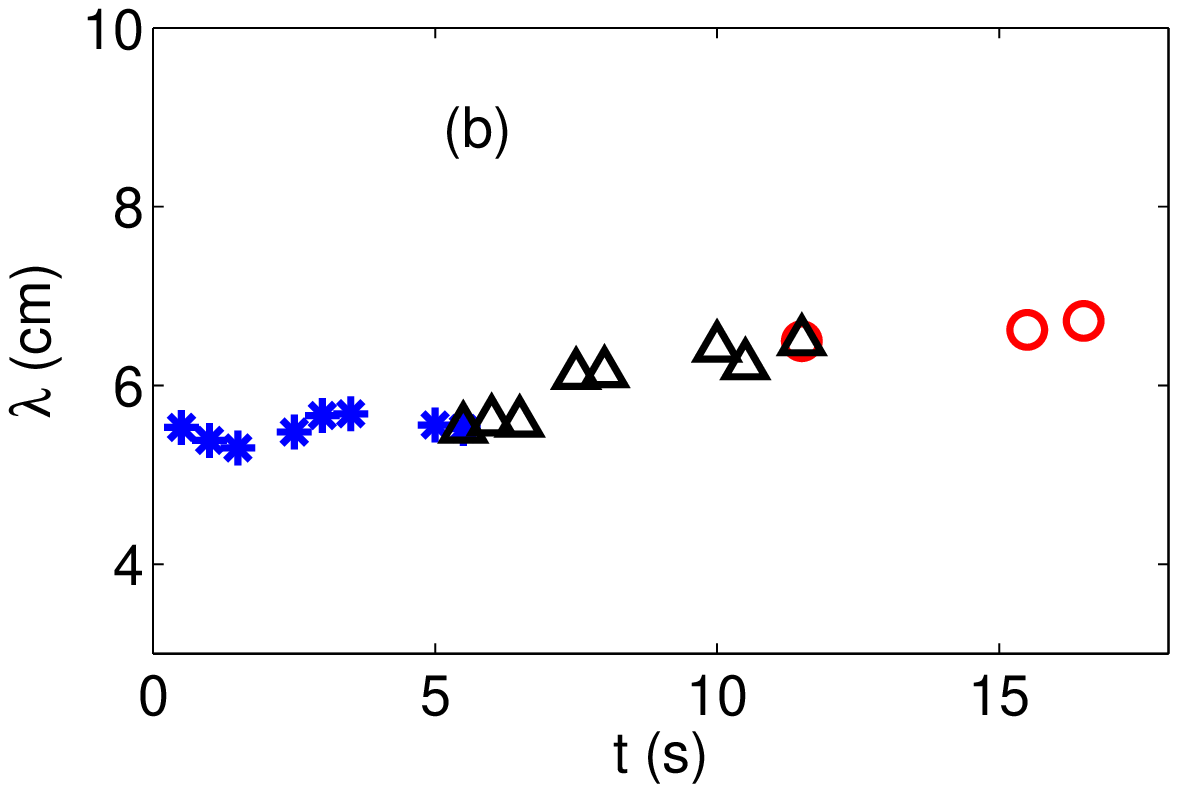}
    \end{center}
   \end{minipage}
\caption{(a) Longitudinal position of a ripple crest $x$ as function of time $t$. (b) Wavelength $\lambda$ of one ripple as a function of time $t$. The symbols correspond to experimental data, and the continuous lines correspond to linear fittings. Asterisks, circles, and triangles correspond to the initial phase of the experiment, to the final phase of the experiment, and to the evolution between the initial and final phases, respectively. The grain diameter was $d_{50}=363\, \mu m$, and the water flow rate was $8.32\,m^3/h$.}
\label{fig:pos_len_300}
\end{figure}

For all test conditions, the numerical code identified, in each frame, the crests of the bedforms and measured their longitudinal positions along the central line of the channel. After that, the wavelengths and celerities were computed. Figure \ref{fig:pos_len_300}a shows the longitudinal position as function of time of a ripple crest when a water flow rate of $8.32\,m^3/h$ was imposed over the $d_{50}=363\, \mu m$ granular bed. The instantaneous celerity of this ripple can be computed from the local slope of the graph presented in the figure. Figure \ref{fig:pos_len_300}a shows a slope variation from the beginning to the end of the experiment, indicating a reduction in the celerity at a final phase. In this figure, the asterisks, circles, and triangles correspond to the initial phase of the experiment, to the final phase of the experiment, and to the evolution between the initial and final phases, respectively. We identified the beginning of the growth of ripples from the initially flat bed as the initial phase. It corresponds to the exponential growth predicted by linear analyses, which is valid for time-scales of the order of the inverse of the growth rate. For larger times, nonlinear analyses predict that ripples continue to grow and eventually saturate while maintaining approximately the same wavelength \cite{Franklin_5}. Concerning the celerity of ripples, it scales with the inverse of their height \cite{Andreotti_1, Elbelrhiti, Franklin_4}, and a decrease in celerity is expected due to the growth of ripples. This is shown in Fig. \ref{fig:pos_len_300}a, where the local slope decreases along time, reaching a final stable slope. Therefore, we identified an initial slope, corresponding to the initial phase, and a final slope, corresponding to the final phase. It should be noted that we are concerned with the initial phases, both linear and nonlinear, of ripple formation and migration without coalescence. Therefore, the final phase corresponds to ripples after the linear phase of instability and before any ripples coalesce.

Figure \ref{fig:pos_len_300}b shows the wavelength as a function of time of one ripple when a flow of $8.32\,m^3/h$ was imposed over the granular bed. The symbols are the same as those in Fig. \ref{fig:pos_len_300}a. Figure \ref{fig:pos_len_300}b shows a slight increase in the wavelength throughout time. This is in accordance with the nonlinear stability analyses that predict a saturation of the amplitude maintaining the same length scale \cite{Franklin_5}.

\begin{figure}
   \begin{minipage}[c]{.49\linewidth}
    \begin{center}
     \includegraphics[width=\linewidth]{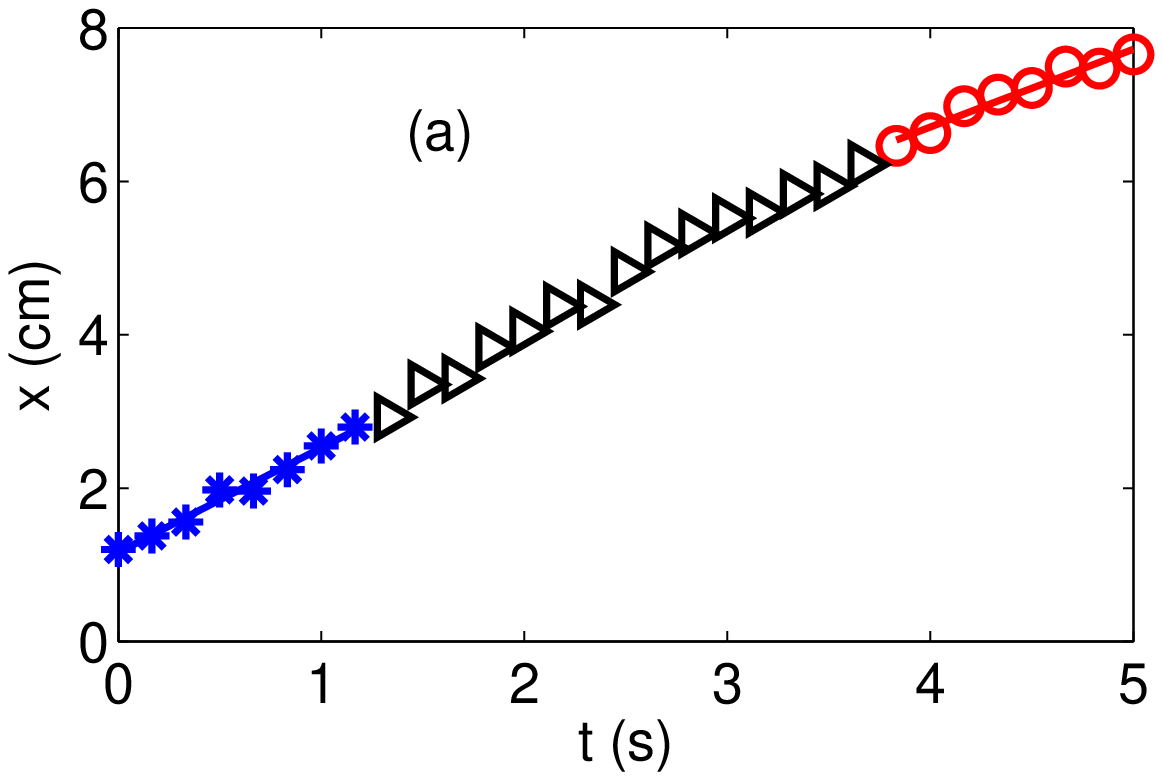}
    \end{center}
   \end{minipage} \hfill
   \begin{minipage}[c]{.49\linewidth}
    \begin{center}
      \includegraphics[width=\linewidth]{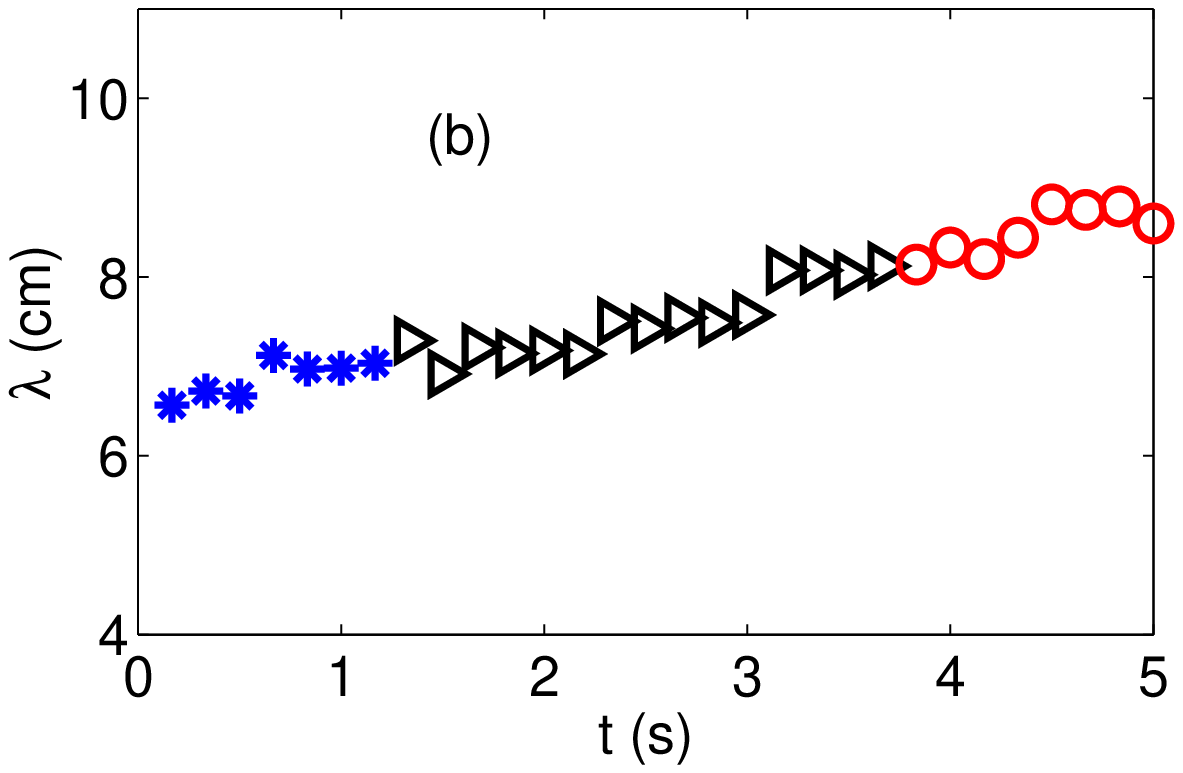}
    \end{center}
   \end{minipage}
\caption{(a) Longitudinal position of a ripple crest $x$ as function of time $t$. (b) Wavelength $\lambda$ of one ripple as a function of time $t$. The symbols correspond to experimental data, and the continuous lines correspond to linear fittings. Asterisks, circles, and triangles correspond to the initial phase of the experiment, to the final phase of the experiment, and to the evolution between the initial and final phases, respectively. The grain diameter was $d_{50}=550\, \mu m$ and the water flow rate was $10.00\,m^3/h$.}
\label{fig:pos_len_500}
\end{figure}

Figure \ref{fig:pos_len_500}a shows the longitudinal position as function of time of a ripple crest, and Fig. \ref{fig:pos_len_500}b shows the wavelength as a function of time of one ripple when a flow of $10.00\,m^3/h$ was imposed over the  $d_{50}=550\, \mu m$ granular bed. The symbols are the same as those in Figs. \ref{fig:pos_len_300}a and \ref{fig:pos_len_300}b. The same behavior, shown in Figs. \ref{fig:pos_len_300}a, \ref{fig:pos_len_300}b, \ref{fig:pos_len_500}a, and \ref{fig:pos_len_500}b, was observed for the other test conditions.

\begin{table}[ht]
\caption{Water flow rate $Q$, Reynolds number $Re$, cross-sectional mean velocity $U$, shear velocity $u_*$, initial wavelength $\lambda_{ini}$, final wavelength $\lambda_{fin}$, initial celerity $c_{ini}$, and final celerity $c_{fin}$ for the $d_{50}=256\, \mu m$ beds.}
\label{table:table1}
\centering
\begin{tabular}{c c c c c c c c}  
\hline\hline
$Q$ & $Re$  & $U$ & $u_*$ & $\lambda_{ini}$ & $\lambda_{fin}$ & $c_{ini}$ & $c_{fin}$ \\
$m^3/h$ & $\cdots$ & $m/s$ & $m/s$ & $m$ & $m$ & $m/s$ & $m/s$ \\ [0.5ex] 
\hline 
6.63 & $2.3\cdot 10^4$ & $0.268$ & 0.0157 & 0.057 & 0.067 & 0.0003 & 0.0002\\
7.12 & $2.5\cdot 10^4$ & $0.288$ & 0.0168 & 0.035 & 0.046 & 0.0011 & 0.0005\\ 
7.38 & $2.6\cdot 10^4$ & $0.300$ & 0.0170 & 0.038 & 0.047 & 0.0014 & 0.0009\\ 
7.72 & $2.7\cdot 10^4$ & $0.312$ & 0.0176 & 0.043 & 0.046 & 0.0017 & 0.0011\\ 
7.99 & $2.8\cdot 10^4$ & $0.323$ & 0.0181 & 0.035 & 0.039 & 0.0023 & 0.0015\\ 
8.31 & $2.9\cdot 10^4$ & $0.336$ & 0.0187 & 0.037 & 0.046 & 0.0039 & 0.0024\\
8.60 & $3.0\cdot 10^4$ & $0.347$ & 0.0192 & 0.038 & 0.043 & 0.0043 & 0.0031\\
8.78 & $3.0\cdot 10^4$ & $0.355$ & 0.0195 & 0.037 & 0.041 & 0.0056 & 0.0031\\
9.15 & $3.2\cdot 10^4$ & $0.369$ & 0.0202 & 0.034 & 0.043 & 0.0075 & 0.0061\\ [1ex] 
\hline 
\end{tabular} 
\end{table}

\begin{table}[ht]
\caption{Water flow rate $Q$, Reynolds number $Re$, cross-sectional mean velocity $U$, shear velocity $u_*$, initial wavelength $\lambda_{ini}$, final wavelength $\lambda_{fin}$, initial celerity $c_{ini}$, and final celerity $c_{fin}$ for the $d_{50}=363\, \mu m$ beds.}
\label{table:table2}
\centering
\begin{tabular}{c c c c c c c c}  
\hline\hline
$Q$ & $Re$  & $U$ & $u_*$ & $\lambda_{ini}$ & $\lambda_{fin}$ & $c_{ini}$ & $c_{fin}$ \\
$m^3/h$ & $\cdots$ & $m/s$ & $m/s$ & $m$ & $m$ & $m/s$ & $m/s$ \\ [0.5ex] 
\hline 
7.15 & $2.5\cdot 10^4$ & $0.288$ & 0.0175 & 0.054 & 0.073 & 0.0002 & 0.0002\\
7.23 & $2.5\cdot 10^4$ & $0.292$ & 0.0176 & 0.056 & 0.065 & 0.0003 & 0.0003\\ 
7.25 & $2.5\cdot 10^4$ & $0.293$ & 0.0177 & 0.064 & 0.069 & 0.0004 & 0.0003\\ 
7.60 & $2.6\cdot 10^4$ & $0.307$ & 0.0185 & 0.058 & 0.062 & 0.0010 & 0.0008\\ 
7.70 & $2.7\cdot 10^4$ & $0.311$ & 0.0188 & 0.061 & 0.067 & 0.0012 & 0.0008\\ 
8.24 & $2.9\cdot 10^4$ & $0.333$ & 0.0200 & 0.061 & 0.067 & 0.0018 & 0.0013\\
8.32 & $2.9\cdot 10^4$ & $0.336$ & 0.0202 & 0.055 & 0.066 & 0.0035 & 0.0020\\
9.24 & $3.2\cdot 10^4$ & $0.373$ & 0.0224 & 0.058 & 0.058 & 0.0059 & 0.0046\\
10.22 & $3.5\cdot 10^4$ & $0.413$ & 0.0247 & 0.054 & 0.061 & 0.0126 & 0.0090\\ [1ex] 
\hline 
\end{tabular} 
\end{table}

\begin{table}[ht]
\caption{Water flow rate $Q$, Reynolds number $Re$, cross-sectional mean velocity $U$, shear velocity $u_*$, initial wavelength $\lambda_{ini}$, final wavelength $\lambda_{fin}$, initial celerity $c_{ini}$, and final celerity $c_{fin}$ for the $d_{50}=550\, \mu m$ beds.}
\label{table:table3}
\centering
\begin{tabular}{c c c c c c c c}  
\hline\hline
$Q$ & $Re$  & $U$ & $u_*$ & $\lambda_{ini}$ & $\lambda_{fin}$ & $c_{ini}$ & $c_{fin}$ \\
$m^3/h$ & $\cdots$ & $m/s$ & $m/s$ & $m$ & $m$ & $m/s$ & $m/s$ \\ [0.5ex] 
\hline 
8.78 & $3.0\cdot 10^4$ & $0.355$ & 0.0213 & 0.067 & 0.081 & 0.0039 & 0.0003\\
9.02 & $3.1\cdot 10^4$ & $0.364$ & 0.0219 & 0.075 & 0.088 & 0.0042 & 0.0011\\ 
9.15 & $3.2\cdot 10^4$ & $0.369$ & 0.0222 & 0.071 & 0.083 & 0.0058 & 0.0039\\
9.70 & $3.4\cdot 10^4$ & $0.391$ & 0.0235 & 0.068 & 0.086 & 0.0093 & 0.0043\\
10.00 & $3.5\cdot 10^4$ & $0.404$ & 0.0242 & 0.068 & 0.085 & 0.0136 & 0.0056\\
10.25& $3.6\cdot 10^4$ & $0.414$ & 0.0247 & 0.076 & 0.084 & 0.0148 & 0.0075\\ [1ex] 
\hline 
\end{tabular} 
\end{table}

\begin{table}[ht]
\caption{Water flow rate $Q$, mean grain diameter $d_{50}$, Shields number $\theta$, particle Reynolds number $Re_*$, initial wavelength normalized by the grain diameter $\lambda_{ini}/d_{50}$, final wavelength normalized by the grain diameter $\lambda_{fin}/d_{50}$, initial celerity normalized by the shear velocity $c_{ini}/u_*$, and final celerity normalized by the shear velocity $c_{fin}/u_*$.}
\label{table:table4}
\centering
\begin{tabular}{c c c c c c c c}  
\hline\hline
$Q$ & $d_{50}$ & $\theta$ & $Re_*$ & $\lambda_{ini}/d_{50}$ & $\lambda_{fin}/d_{50}$ & $c_{ini}/u_*$ & $c_{fin}/u_*$ \\
$m^3/h$ & $\mu m$ & $\cdots$ & $\cdots$ & $\cdots$ & $\cdots$ & $\cdots$ & $\cdots$ \\ [0.5ex] 
\hline 
6.63 & $256$ & 0.065 & 4 & 222 & 263 & 0.0188 & 0.0144\\
7.12 & $256$ & 0.073 & 4 & 137 & 181 & 0.0655 & 0.0296\\ 
7.38 & $256$ & 0.077 & 4 & 149 & 184 & 0.0832 & 0.0551\\ 
7.72 & $256$ & 0.082 & 5 & 168 & 181 & 0.0956 & 0.0632\\ 
7.99 & $256$ & 0.087 & 5 & 138 & 152 & 0.1272 & 0.0827\\ 
8.31 & $256$ & 0.093 & 5 & 143 & 178 & 0.2102 & 0.1303\\
8.60 & $256$ & 0.098 & 5 & 148 & 168 & 0.2233 & 0.1592\\
8.78 & $256$ & 0.101 & 5 & 145 & 161 & 0.2877 & 0.1605\\
9.15 & $256$ & 0.108 & 5 & 132 & 168 & 0.3694 & 0.3042\\
7.15 & $363$ & 0.057 & 6 & 175 & 201 & 0.0134 & 0.0119\\
7.23 & $363$ & 0.058 & 6 & 155 & 179 & 0.0194 & 0.0145\\ 
7.25 & $363$ & 0.059 & 6 & 176 & 189 & 0.0246 & 0.0171\\ 
7.60 & $363$ & 0.064 & 7 & 160 & 170 & 0.0546 & 0.0432\\ 
7.70 & $363$ & 0.066 & 7 & 167 & 186 & 0.0620 & 0.0431\\ 
8.24 & $363$ & 0.075 & 7 & 168 & 185 & 0.0909 & 0.0646\\
8.32 & $363$ & 0.076 & 7 & 152 & 182 & 0.1722 & 0.0986\\
9.24 & $363$ & 0.094 & 8 & 160 & 158 & 0.2628 & 0.2037\\
10.22 & $363$ & 0.114 & 9 & 150 & 168 & 0.5096 & 0.3628\\
8.78 & $550$ & 0.056 & 12 & 122 & 148 & 0.1819 & 0.1771\\
9.02 & $550$ & 0.059 & 12 & 136 & 160 & 0.1918 & 0.1784\\ 
9.15 & $550$ & 0.061 & 12 & 129 & 151 & 0.2606 & 0.2371\\ 
9.70 & $550$ & 0.068 & 13 & 124 & 156 & 0.3977 & 0.2812\\
10.00 & $550$ & 0.072 & 13 & 124 & 155 & 0.5632 & 0.4186\\
10.25 & $550$ & 0.075 & 14 & 138 & 153 & 0.5984 & 0.4406\\ [1ex] 
\hline 
\end{tabular} 
\end{table} 

On the basis of the dependencies of the crest positions and measured wavelengths on time, the mean celerities and wavelengths were computed for each test condition at initial and final phases. For each water flow rate, Tables \ref{table:table1}, \ref{table:table2}, and \ref{table:table3} present the Reynolds number $Re$, the cross-sectional mean velocity $U$, the shear velocity $u_*$, the initial wavelength $\lambda_{ini}$, the final wavelength $\lambda_{fin}$, the initial celerity $c_{ini}$, and the final celerity $c_{fin}$. Table \ref{table:table1} concerns the $d_{50}=256\, \mu m$ beds, Tab. \ref{table:table2} concerns the $d_{50}=363\, \mu m$ beds, and Tab. \ref{table:table3} concerns the $d_{50}=550\, \mu m$ beds.

We note from Tabs. \ref{table:table1} to \ref{table:table3} that the celerity of the ripples increases with $u_*$. This was expected because the celerity of the ripples and dunes scales with the bed-load transport rate $Q_B$ and the inverse of their height $h$ \cite{Andreotti_1, Elbelrhiti, Franklin_4}:

\begin{equation}
c\,\sim\,\frac{Q_B}{h}
\label{scaling_celerity}
\end{equation}

One of the most used expressions for the transport rate is that of Meyer-Peter and M\"{u}ller (1948) \cite{Mueller}, given by

\begin{equation}
\phi_B\,=\,8(\theta-\theta_{th})^{3/2}
\label{exp_transp_rate}
\end{equation}

\noindent In Eq. \ref{exp_transp_rate}, $\phi_B\,=\,q_B((S-1)gd^3)^{-1/2}$ is the normalized volumetric bed-load transport rate, where $S=\rho_s / \rho$ is the ratio between the specific masses of the grains $\rho_s$ and the fluid $\rho$; $g$ is the gravitational acceleration; and $q_B$ is the volumetric bed-load transport rate per unit width. The Shields number $\theta$ is defined as the ratio between the shear force caused by the fluid and the grains' weight

\begin{equation}
\theta\, =\, \frac{\tau}{(\rho_{s}-\rho)gd}
\label{shields}
\end{equation}

\noindent where $\tau$ is the shear stress caused by the fluid on the bed. In the case of turbulent boundary layers, $\tau\,=\,\rho u_*^2$. Finally, $\theta_{th}$ is the Shields number corresponding to the threshold shear stress for bed-load incipient motion \cite{Raudkivi_1}. From Eqs. \ref{exp_transp_rate} and \ref{shields}, the celerity of the ripples scales with $u_*^3$.

From the experimental data, the wavelength of the ripples shows no clear variation with $u_*$, but it varies with $d_{50}$. This is not in agreement with some stability analyses and experiments for ripples in liquid flows \cite{Charru_3, Franklin_4, Kuru, Franklin_3, Pahtz}. However, this agrees with the experiments of Coleman et al. (2003) \cite{Coleman_1}, who found that the wavelength of the ripples does not vary with the water flow conditions and varies with the grain diameter.

To determine the bed-load transport rate and to analyze the formation of ripples, some dimensionless parameters are necessary. They are usually taken as $\theta$, defined in Eq. \ref{shields}, and the particle Reynolds number, $Re_*$. $Re_*$ is the Reynolds number at the grain scale \cite{Raudkivi_1}

\begin{equation}
Re_* = \frac{u_*d}{\nu}
\label{particle_rey}
\end{equation}

\noindent where $\nu$ is the kinematic viscosity of the fluid. It can also be seen as the ratio between $d$ and the viscous length $\nu/u_*$, giving an indication if the fluid flow occurs in the hydraulic smooth regime $Re_*<5$ or in the hydraulic rough regime $Re_*>70$ \cite{Schlichting_1}. In the present study, we compute $Re_*$ using $d_{50}$.

For each water flow rate and grain diameter, Tab. \ref{table:table4} presents the Shields number $\theta$, the particle Reynolds number $Re_*$, the initial wavelength normalized by the grain diameter $\lambda_{ini}/d_{50}$, the final wavelength normalized by the grain diameter $\lambda_{fin}/d_{50}$, the initial celerity normalized by the shear velocity $c_{ini}/u_*$, and the final celerity normalized by the shear velocity $c_{fin}/u_*$.

The experimental data do not show a clear variation of the dimensionless ripples wavelength with the fluid flow conditions. For the majority of tested conditions, the ripple wavelength is within $100\,<\lambda_{ini}/d_{50}\,<\,200$ and $100\,<\lambda_{fin}/d_{50}\,<\,200$. This range is in good agreement with the experimental study of Coleman et al. (2003) \cite{Coleman_1} and with some open-channel experiments, such as Coleman and Melville (1996) \cite{Coleman}. Coleman et al. (2003) \cite{Coleman_1} found that ripples in closed-conduit flows and in subcritical open-channel flows have roughly the same wavelength, and proposed the following correlation for the wavelength of the initial ripples

\begin{equation}
\lambda_{ini}\,=\,175d^{0.75}
\label{coleman}
\end{equation}

\noindent However, the present experimental data do not fit Eq. \ref{coleman}. As noted by Franklin (2010) \cite{Franklin_4}, Eq. \ref{coleman} has a dimensional inconsistency. In this equation, $\lambda_{ini}$ is proportional to the diameter raised to a power of $0.75$, so that the numerical constant has a dimension. Therefore, it is difficult to imagine Eq. \ref{coleman} as a universal expression.

We note that finite size effects could influence the correlation between the wavelength and $u_*$. Although the liquid flow occurs in a cross section that is 160 mm wide by 43 mm high, the lateral walls affect the shear velocity in the central region of the channel: as the water flow rate is increased, velocity gradients in the span direction get higher close to the vertical walls, so that the span distribution of velocities gets flattened around the central line of the channel. As a consequence, the shear velocities in the central region, where the wavelengths were measured, may slightly differ from the estimated values.

\begin{figure}
   \begin{minipage}[c]{.49\linewidth}
    \begin{center}
     \includegraphics[width=\linewidth]{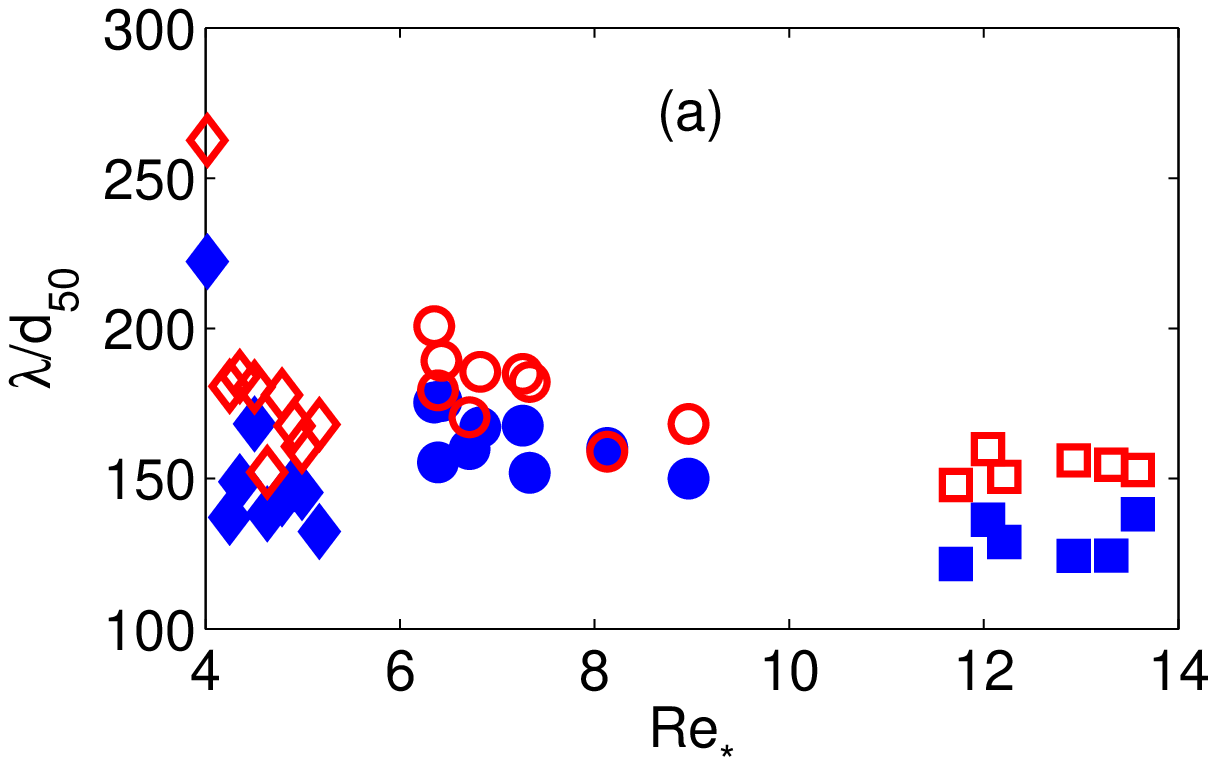}
    \end{center}
   \end{minipage} \hfill
   \begin{minipage}[c]{.49\linewidth}
    \begin{center}
      \includegraphics[width=\linewidth]{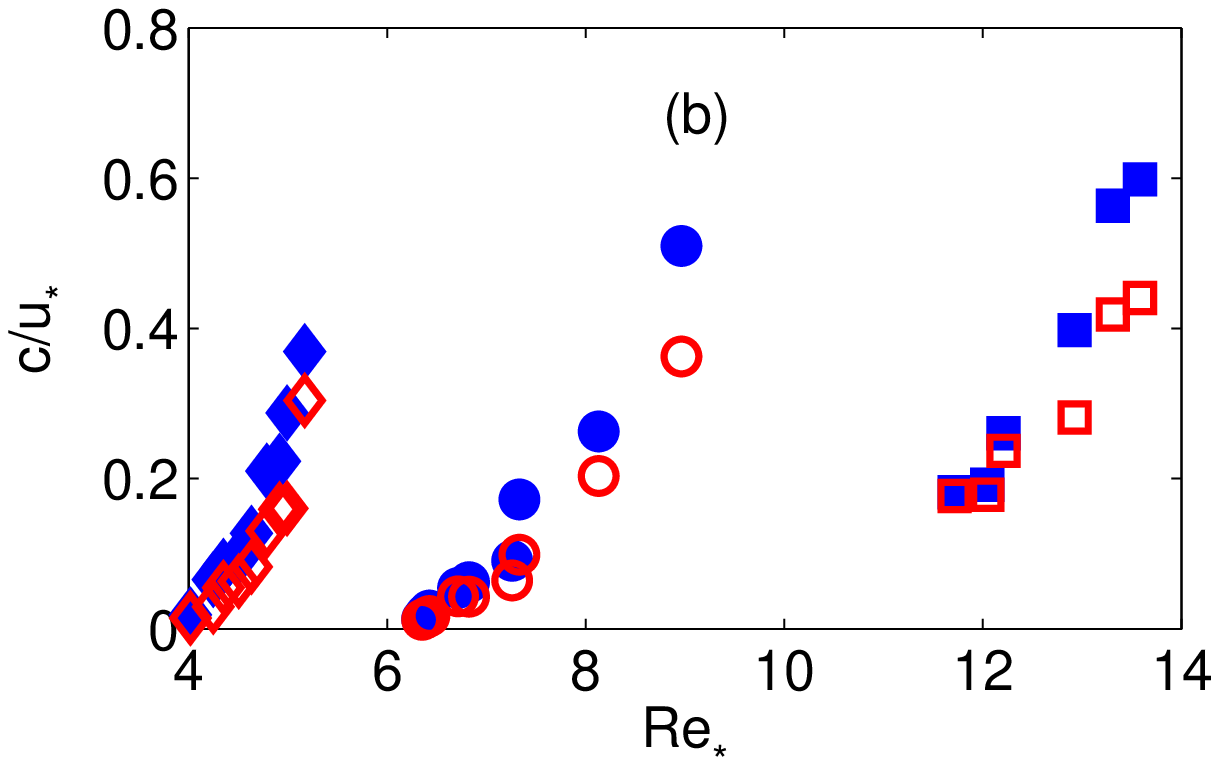}
    \end{center}
   \end{minipage}
\caption{(a) Wavelength normalized by the grain diameter $\lambda/d_{50}$ as function of the particle Reynolds number $Re_*$. (b) Celerity normalized by the shear velocity $c/u_*$ as function of $Re_*$. Diamonds, circles, and squares correspond to the $d_{50}=256\, \mu m$, $d_{50}=363\, \mu m$, and $d_{50}=550\, \mu m$ beds, respectively. Filled symbols correspond to the initial phase, and open symbols correspond to the final phase.}
\label{fig:L_c_Rep}
\end{figure}

Figure \ref{fig:L_c_Rep}a presents the initial and final wavelengths normalized by the grain diameters $\lambda/d_{50}$ as a function of $Re_*$. The diamonds, circles, and squares correspond to the $d_{50}=256\, \mu m$, $d_{50}=363\, \mu m$, and $d_{50}=550\, \mu m$ beds, respectively, and the filled and open symbols correspond to the initial and final phases, respectively. No clear variation in the dimensionless wavelength with the fluid flow conditions can be observed from this figure.

Figure \ref{fig:L_c_Rep}b presents the ripple celerity normalized by the shear velocity $c/u_*$ as function of $Re_*$. The symbols are the same as in Fig. \ref{fig:L_c_Rep}a. Figure \ref{fig:L_c_Rep}b shows that the dimensionless celerity of the ripples increases with the water flow conditions.

Because there is a threshold below which the granular bed is static and above which the bed load takes place, the value of the threshold must be considered in a scaling law between the celerity and the fluid flow conditions. From Eq. \ref{exp_transp_rate} and considering that the ripple celerity varies with $\phi_B$, we can expect to scale $c/u_*$ with $\theta-\theta_{th}$. The difficulty here is knowledge of the correct value of $\theta_{th}$. There is a large dispersion of the experimental $\theta_{th}$ values in the Shields diagram \cite{Raudkivi_1}. In addition, Charru et al. (2004) \cite{Charru_1} showed that the surface density of moving grains decays with time while their velocity remains unchanged. They proposed that this decay is due to an increase in bed compactness caused by the rearrangement of grains, known as armoring, which leads to an increase in the threshold shear rate for the bed load. We did not measure the threshold Shields number because the obtained values would be subject to large deviations due to bed armoring. On the basis of the Shields diagram, we employed $\theta_{th}\,=\,0.047$ for the $d_{50}=256\, \mu m$ and $d_{50}=363\, \mu m$ grains and $\theta_{th}\,=\,0.015$ for the $d_{50}=550\, \mu m$ grains. In the present experiments, the basic flow was in hydraulic smooth regime (or in the beginning of the smooth-rough transition) \cite{Franklin_8}, meaning that, for the same water velocity profiles, the exposed grains are subject to the same velocity gradients; therefore, larger grains tend to be exposed to larger fluid velocities. Consequently, we inferred that the largest grains are expected to have lower threshold Shields numbers.

\begin{figure}
   \begin{minipage}[c]{.49\linewidth}
    \begin{center}
     \includegraphics[width=\linewidth]{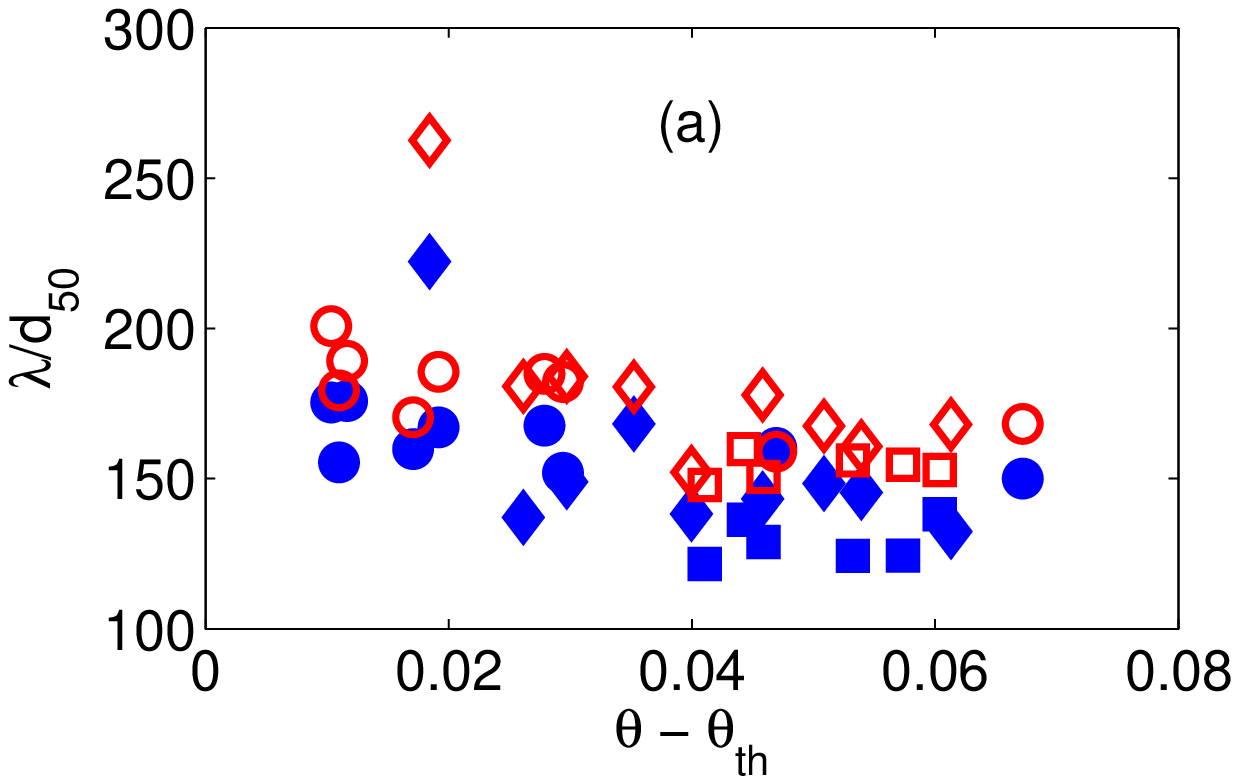}
    \end{center}
   \end{minipage} \hfill
   \begin{minipage}[c]{.49\linewidth}
    \begin{center}
      \includegraphics[width=\linewidth]{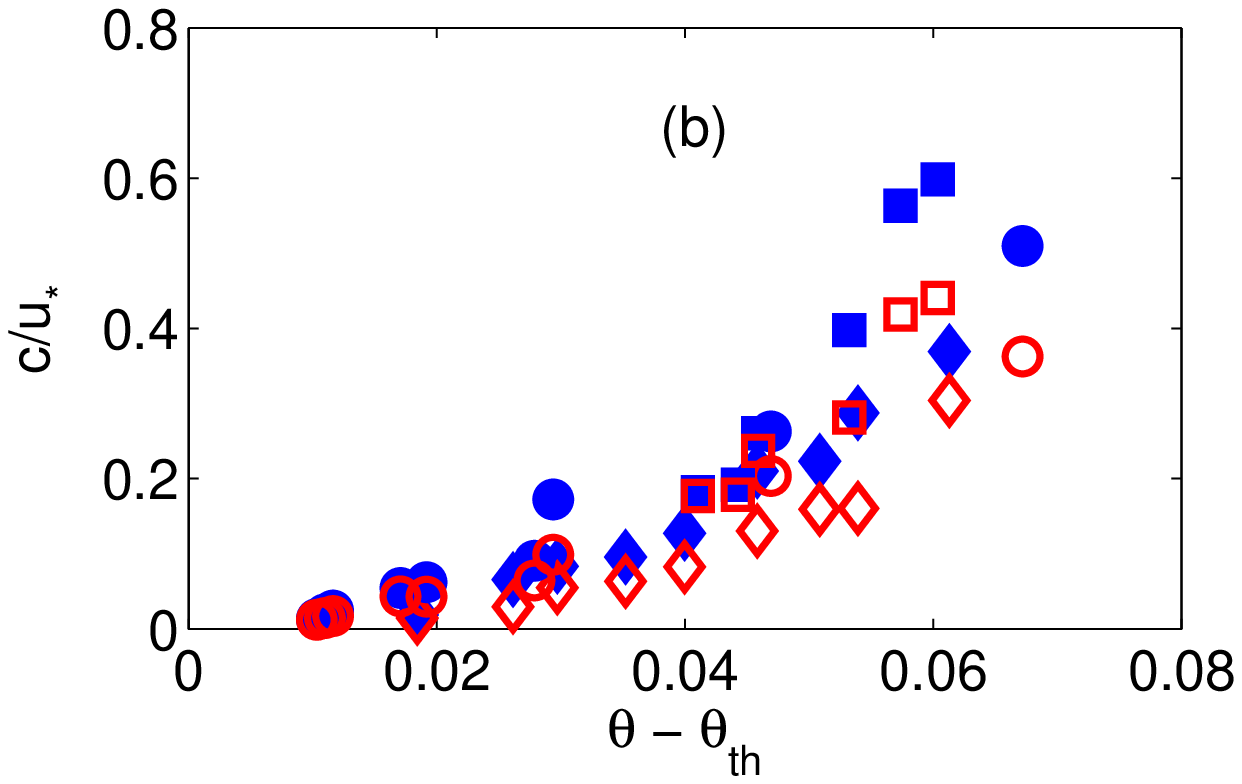}
    \end{center}
   \end{minipage}
\caption{(a) Wavelength normalized by the grain diameter $\lambda/d_{50}$ as a function of the difference between the Shields number and its threshold value $\theta - \theta_{th}$. (b) Celerity normalized by the shear velocity $c/u_*$ as a function of the difference between the Shields number and its threshold value $\theta - \theta_{th}$. Diamonds, circles, and squares correspond to the $d_{50}=256\, \mu m$, $d_{50}=363\, \mu m$, and $d_{50}=550\, \mu m$ beds, respectively. Filled symbols correspond to the initial phase, and open symbols correspond to the final phase.}
\label{fig:L_c_theta_thetath}
\end{figure}

Figures \ref{fig:L_c_theta_thetath}a and \ref{fig:L_c_theta_thetath}b present the wavelength normalized by the grain diameter $\lambda/d_{50}$ and the celerity normalized by the shear velocity $c/u_*$, respectively, as functions of the difference between the Shields number and its threshold value $\theta - \theta_{th}$. The symbols are the same as in Fig. \ref{fig:L_c_Rep}. Filled symbols correspond to the initial phase, and open symbols correspond to the final phase. For the  wavelength, as expected from the previous discussion, there is no clear variation with respect to $\theta - \theta_{th}$.

\begin{figure}[!ht]
  \begin{center}
    \includegraphics[width=0.49\columnwidth]{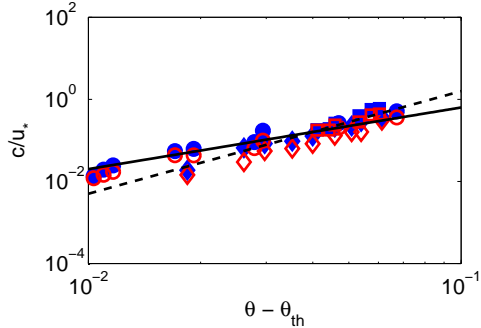}
    \caption{Celerity normalized by the shear velocity $c/u_*$ as a function of the difference between the Shields number and its threshold value $\theta - \theta_{th}$ on a log-log plot. Diamonds, circles, and squares correspond to the $d_{50}=256\, \mu m$, $d_{50}=363\, \mu m$, and $d_{50}=550\, \mu m$ beds, respectively, the continuous line is a $(\theta - \theta_{th})^{3/2}$ function \cite{Mueller}, and the dashed line is a $(\theta - \theta_{th})^{5/2}$ function \cite{Franklin_9}. Filled symbols correspond to the initial phase, and open symbols correspond to the final phase.}
    \label{loglog_c_theta}
  \end{center}
\end{figure}

Concerning the dimensionless celerity, all of the data for the initial and final phases collapse, and a scaling law as a function of $\theta - \theta_{th}$ seems to exist. In order to verify if the experimental data follow a power-law dependence on $\theta - \theta_{th}$, as predicted by many bed-load transport rate equations, the experimental data was plotted on a logarithmic scale. Figure \ref{loglog_c_theta} presents the ripple celerity normalized by the shear velocity $c/u_*$ as function of the difference between the Shields number and its threshold value $\theta - \theta_{th}$ on a log-log plot. The symbols are the same as those in Figs. \ref{fig:L_c_Rep} and \ref{fig:L_c_theta_thetath}, the continuous line is a $(\theta - \theta_{th})^{3/2}$ function, and the dashed line is a $(\theta - \theta_{th})^{5/2}$ function. The two lines are included as references and correspond to two proposed transport rate equations: the Meyer-Peter and  M\"{u}ller (1948) equation \cite{Mueller}, given by Eq. \ref{exp_transp_rate}, and an equation proposed by Franklin and Charru (2011) \cite{Franklin_9}, given by 

\begin{equation}
\phi_B\,=\,34Re_s \left(\theta - \theta_{th} \right )^{2.5}
\label{franklin}
\end{equation}

\noindent where $Re_s\,=\,U_sd/\nu$ is the Reynolds number based on the settling velocity of a single grain, $U_s$.

The totality of the experimental data fits the $(\theta - \theta_{th})^{3/2}$ function better, such that $(\theta - \theta_{th})^{3/2}$ is a better fitting for ripple celerity. This shows that the Meyer-Peter and  M\"{u}ller (1948) \cite{Mueller} equation can be used for the scaling between the celerity and the bed-load transport rate (Eq. \ref{scaling_celerity}).

Far from the threshold in the $( \theta - \theta_{th} )\,>\, 3\cdot 10^{-2}$  region, the experimental data fit the $(\theta - \theta_{th})^{5/2}$ function better, indicating that the Franklin and Charru (2011) \cite{Franklin_9} equation can be used in this region. This can be explained by the fact that the Meyer-Peter and  M\"{u}ller (1948) \cite{Mueller} equation is based on gravitational flow experiments with grain diameters in the range of $0.4\,mm\, \leq\,d\, \leq\,30\,mm$, whereas the Franklin and Charru (2011) \cite{Franklin_9} equation is based on pressure-driven flow experiments with grain diameters in the range of $0.1\,mm\, <\,d\, <\, 0.5\,mm$, which is similar to the present experiments. The larger grains used in \cite{Mueller} tend to be associated with Shields numbers closer to the threshold values.

\section{Conclusion}

This paper presented an experimental study on the formation and migration of ripples in a turbulent liquid flow. Turbulent water flows were imposed over initially flat granular beds in a closed conduit, and their evolutions were filmed with a high-definition camera. A numerical code was developed to treat the images on the basis of the identification of crests and troughs from gray-level differences. The code is able to determine the wavelengths and celerities from the acquired images.

For each flow condition, the experimental results showed that the celerity decreases with time, whereas the wavelength increases slightly. The decrease in celerity is due to the initial growth of ripples because their celerity scales with the inverse of their height. The slight variation in wavelength is in accordance with nonlinear stability analyses that predict a saturation of the amplitude while maintaining the same length scale.

The experiments showed that the celerity of the ripples increases with the the fluid flow conditions. The celerity of the ripples and dunes scales with the bed-load transport rate, which in turn varies with the fluid shear stresses; therefore, an increase in celerity is expected. Because there is a threshold for the bed load, a scaling law between the celerity and the fluid flow condition must consider the threshold value. The totality of experimental data fits the $(\theta - \theta_{th})^{3/2}$ function better, showing that the Meyer-Peter and  M\"{u}ller (1948) \cite{Mueller} equation can be used in the scaling between the celerity and the bed-load transport rate. However, far from the threshold, the experimental data fits the $(\theta - \theta_{th})^{5/2}$ function better, indicating the use of the Franklin and Charru (2011) \cite{Franklin_9} equation in this region.

The experimental data did not show a clear variation in the ripple wavelength with the fluid flow conditions. For the majority of tested conditions, the ripple wavelength was within $100\,<\lambda_{ini}/d_{50}\,<\,200$ and $100\,<\lambda_{fin}/d_{50}\,<\,200$. This is not in agreement with some stability analyses and experiments on ripples in liquid flows \cite{Charru_3, Franklin_4, Kuru, Franklin_3}; however, this agrees with the experiments of Coleman et al. (2003) \cite{Coleman_1}, and with some open-channel experiments, such as Coleman and Melville (1996) \cite{Coleman}. The scaling laws for the wavelength of aquatic ripples in closed conduits are still an open problem, and the present experimental data bring some light to this intricate problem.

\section{Acknowledgments}

\begin{sloppypar}
The authors are grateful to FAPESP (grant no. 2012/19562-6), CNPq (grant no. 471391/2013-1), and FAEPEX/UNICAMP (conv. 519.292, projects 1435/12 and AP0008/2013) for the provided financial support. 
\end{sloppypar}